\documentclass[manuscript, screen]{acmart}

\AtBeginDocument{%
  \providecommand\BibTeX{{%
    \normalfont B\kern-0.5em{\scshape i\kern-0.25em b}\kern-0.8em\TeX}}}

\setcopyright{acmcopyright}
\copyrightyear{2023}
\acmYear{2023}
\acmConference[Submitted to ACM FAccT 2023]{ACM Conference on Fairness, Accountability, and Transparency}{June 12--15, 2023}{Chicago, IL, USA}
\acmDOI{XXXXXXX.XXXXXXX}

\usepackage{enumitem}

\begin{document}

\title[Explainability in AI Policies]{Explainability in AI Policies:  A Critical Review of Communications, Reports, Regulations, and Standards in the EU, US, and UK}

\author{Luca Nannini}
\email{l.nannini@usc.es, lnannini@minsait.com}
\orcid{0000-0002-4733-9760}
\affiliation{
  \institution{{Minsait by Indra Sistemas SA}, {Centro Singular de Investigación en Tecnoloxías Intelixentes (CiTIUS) Universidade de Santiago de Compostela}}
  \streetaddress{Rúa de Jenaro de la Fuente, s/n}
  \city{Santiago de Compostela}
  \state{A Coruña}
  \country{Spain}
  \postcode{15705}
}

\author{Agathe Balayn}
\email{a.m.a.balayn@tudelft.nl}
\orcid{0000-0003-2725-5305}
\affiliation{%
  \institution{Delft University of Technology}
  \country{Netherlands}
}

\author{Adam Leon Smith}
\email{adamleonsmith@protonmail.com}
\orcid{0000-0002-6614-0888}
\affiliation{%
  \institution{Dragonfly}
  \streetaddress{Avinguda Diagonal, 211}
  \city{Barcelona}
  \state{Catalonia}
  \country{Spain}
  \postcode{08018}
}

\renewcommand{\shortauthors}{Nannini et al.}

\begin{abstract}
Public attention towards explainability of artificial intelligence (AI) systems has been rising  in recent years to offer methodologies for human oversight. This has translated into the proliferation of research outputs, such as from Explainable AI, to enhance transparency and control for system debugging and monitoring, and intelligibility of system process and output for user services. Yet, such outputs are difficult to adopt on a practical level due to a lack of a common regulatory baseline, and the contextual nature of explanations. Governmental policies are now attempting to tackle such exigence,  however it remains unclear to what extent published communications, regulations, and standards adopt an informed perspective to support research, industry, and civil interests. In this study, we perform the first thematic and gap analysis of this plethora of policies and standards on explainability in the EU, US, and UK. Through a rigorous survey of policy documents, we first contribute an overview of governmental regulatory trajectories within AI explainability and its sociotechnical impacts.
We find that policies are often informed by coarse notions and requirements for explanations. This might be due to the willingness to conciliate explanations foremost as a risk management tool for AI oversight, but also due to the lack of a consensus on what constitutes a valid algorithmic explanation, and how feasible the implementation and deployment of such explanations are across stakeholders of an organization.
Informed by AI explainability research, we conduct a gap analysis of existing policies, leading us to formulate a set of recommendations on how to address explainability in regulations for AI systems, especially discussing the definition, feasibility, and usability of explanations, as well as allocating accountability to explanation providers.
\end{abstract}




\maketitle

\section{Introduction}
The ability of artificial intelligence (AI) systems to explain their decision-making processes, also known as "\textit{explainability}", has become an increasingly important AI governance topic to ensure that systems are transparent and accountable \cite{stanford}. Governments are now pacing up strategies to foster AI innovation while mitigating potential risks through explanations. 
On the other side, AI explainability has also become a prominent research focus in  the machine learning and human-computer interaction communities 
\cite{holzinger2018current,abdul2018trends}. The research outputs around AI explainability oftentimes reveal the importance of the context of use, for which legislation might miss a clear and informed baseline to implement explainable AI methods appropriately. Hence, facing the plethora of recent policy documents, and the known complexity of the concept of AI explainability, one needs to understand to what extent the documents are informed by the research and are cognizant and account for  current explainable AI challenges, so as to design policy documents that further foster a responsible use of AI in the future. 

To shed light over the current state of policies around AI explainability and inform future regulatory trajectories, 
we conduct a rigorous, comprehensive analysis of existing governmental policies in the European Union (EU), the United States of America (US), and the United Kingdom (UK). In terms of official governmental communications, we consider commissioned white papers and guidelines, as well as bills and standardization documents impacting the fruition of AI explanations. Next, we identify key requirements for explainability in policy documents, considering aspects such as transparency, interpretability, and explainability in decision-making.
We then evaluate the suitability of each governmental policy through a gap analysis informed by  research publications on AI explainability,
and develop recommendations for addressing these issues. 

Despite the diversity of policy documents, we observe primarily an urge to enact strategic plans for AI research and development (R\&D) and a multifaceted consideration of
explainability as a tool to downside innovation risks and harms within civil rights.
We also find that these documents do not necessarily account for the complexity and recency of the concept of explainability in AI, especially in terms of 
definition, feasibility, and usability of AI explanations. 
This leads us to reflect upon the discretion of explanation providers, also connected to how ambiguities in policy terminology and procedures might further lead to implementation challenges and failures, and ethics-washing opportunities. 
Ultimately, our study provides a valuable starting point for understanding the current state of AI explainability in policies and standards, and  finally brings together an unified view over regulatory approaches to AI explainability. It bears implications for policy-makers but also for AI researchers, calling for documents and research that would account for identified limitations in this ever-evolving field, while being specific enough to avoid ethics-washing.

The rest of the manuscript is organized as follows. 
Section \hyperref[sec:2]{2} provides background knowledge on the rising attention towards AI explainability,  from the lens of both 
technical Explainable AI approaches, and more comprehensive sociotechnical implementations.
Section \hyperref[sec:3]{3} details the mixed-method approach we followed for this study. 
Section \hyperref[sec:4]{4} maps explainability regulations and standards across countries, 
performing a gap analysis as informed by potential intergovernmental discrepancies and constraints. 
Section \hyperref[sec:5]{5} analyses discrepancies 
around the mapped dimensions of explainability documents, reasoning over their actuality in lights of related research publications. 
Finally, section \hyperref[sec:6]{6} and section \hyperref[sec:7]{7} reflects upon current research constraints, findings, and future research and policy directions.



\section{Background: translating AI explainability into regulations}
\label{sec:2}
\subsection{Defining explainability and interpretability in AI research}
Despite the relatively common use of the terms, there is no consensus on the definitions of \emph{explainability} or \emph{interpretability} in the AI standards community.
The ambiguities with the definitions encompass the type of explanation, the reason for the explanation, the purpose of the explanation, and to whom the explanation is provided \cite{arrieta_explainable_2019}.
For the sake of clarity, we intend by \textit{explainability} the communicative ability to deliver insightful information regarding the inner functioning of complex algorithmic architectures.
The definitions of interpretability  \cite{chakraborty_interpretability_2017}
revolve around 
either systems that are designed to be inherently interpretable \cite{rudin_2019_stop}, or  model-agnostic approaches that are based on observing systems' inputs and outputs \cite{ribeiro_model-agnostic_2016}.
By \textit{interpretability}, we refer to the human cognitive ability to inherently understand the functioning of an AI system, intended as the inner relations between its data inputs and its learned output computation functions. 
Once algorithmic architecture scales up, e.g., in terms of hyperparameters and features, 
interrelating complexity arises, jeopardizing direct interpretation. 
We detail below further distinctions as from multiple research domains for explainable AI systems.


Machine learning researchers have primarily focused on developing methods that allow to make the decision process of a machine learning (ML) model, going from an input data sample to the output label produced by the model, less of a black-box. These methods are often categorized along the following dimensions \cite{confalonieri2021historical,das2020opportunities}: the type of data (e.g., tabular \cite{burkart2021survey}, textual \cite{danilevsky2020survey}, visual \cite{zablocki2022explainability}), task (e.g., classification \cite{samek2019towards}, regression \cite{letzgus2022toward}, recommendation \cite{zhang2020explainable}), and algorithm (e.g., different types of ML algorithms or deep learning (DL) architectures) the explanation applies to \cite{belle2021principles}; the nature of the explanation, especially the literature talks either about developing models that are inherently explainable \cite{zhang2018interpretable} or methods that explain black-box models in a post-hoc manner \cite{jesus2021can}; family, as the modes of explanations might vary from reflecting associations between the input and output, to contrasting different input samples and their explanations, or to displaying causal explanations, e.g., through counterfactuals \cite{sokol2020explainability}; the scope of the explanations, especially literature often discusses local explanations about a single input sample or global explanations that refer to the overall model behavior \cite{sokol2020explainability}. 
Challenges in this research area especially revolve around developing faithful explanations for a broad set of ML models \cite{jin2022evaluating}, ensuring these explanations are of high-fidelity (i.e., accurately reflect the model behavior) \cite{balagopalan2022road,sixt2020explanations,adebayo2022post}, and designing usable benchmarks to properly assess the proposed explanations \cite{yang2019bim} (difficulties arise in 
formulating the main properties a \emph{good} explanation should fulfill \cite{lage2019human}).

Human Computer Interaction (HCI) researchers have also investigated what makes a \emph{good} explanation \cite{sokol2020explainability,mohseni2021multidisciplinary}, and how are explanations used in practice \cite{balayn2022can}. Besides the dimensions of explanations highlighted above, they have found additional dimensions that should be accounted for when developing or selecting explainability methods. Especially, the purposes of the explanations for each stakeholder within the AI lifecycle vary (e.g., a decision-subject might need explanations for contestability and recourse, a developer for model debugging, an auditor for assessing the model readiness for deployment), and would value best different types of explanation typologies \cite{sokol2020explainability}. The usability of the explanation \cite{sokol2020explainability}  also varies across stakeholders, based e.g., on the explanations' complexity, completeness, interactiveness, or medium (e.g., visual or textual mode of explanation), etc. 
Studies with various stakeholders have been performed to understand their use of explanations in given scenarios or in their daily practices. Next to the potential usefulness of the explanations, these studies identified disparate use of a narrow set of explanations in practice, and various types of biases hindering a trustworthy use of the explanations (e.g., misinterpretations, confirmation bias, lack of critical attitude towards presented numbers, etc.) \cite{balayn2022can,alqaraawi2020evaluating,ehsan2021explainable}.
The challenge hence remains to make existing explanations appropriately usable by stakeholders with different backgrounds \cite{ehsan2020human,hadash2022improving,szymanski2021visual}, as well as to better understand their needs and how new explanations should be designed to account for these needs \cite{liao2020questioning}.


\subsection{Policy versus research}

Explainability in AI can be considered a multidisciplinary field that involves various aspects such as technical, ethical, legal, and social \cite{hackerVarietiesAIExplanations2022}.
The variety of XAI methods and purposes proposed in research reflects the diversity of categorical definitions for AI systems \cite{russel2013artificial}. As a result, ambiguities pose risks to the public perception of these systems' capabilities and limitations \cite{johnsonverdicchio, kelleyetalexciting}. This might contribute to confusion and decreased awareness over how such systems are human-produced artifacts, reflecting prior sociotechnical instances and choices informing their design \cite{johnsonverdicchio, Elishanddanahboyd}. This discrepancy is mirrored in how AI system developers, such as ML engineers, tend to differently define these systems and their capabilities in contrast to policymakers \cite{krafftetaldefining}.
Such a discrepancy hence poses risks within the debate on regulating AI systems. This is indeed now signaled by the increasing attention within the field of AI ethics to operationalize human principles \cite{morley2021operationalising, ibanez}, ascribable to the definition of \textit{Second-wave of AI Ethics} \cite{hickok2021lessons}. This research attention aims to operationalize AI principles through a sociotechnical top-down approach to AI governance in organizations \cite{mittelstadt2019principles, ayling2022putting}. In fact, only claiming to adhere to AI ethics principles might translate into ethics-washing practices
if policies, such as accountability measures, are not implemented within a clear regulatory baseline and legal recourse mechanisms \cite{Floridi2019, bietti2020ethics}. 
For such reasons, establishing a proactive regulatory approach to define and regulate explainability of AI systems is crucial to ensure their ethical alignment with human values and principles \cite{Fenwick2018}. Parallel to this second-wave, scholars are now inquiring over the suitability of explainability in governmental regulatory initiatives such as bills and enacted laws. Among the most relevant debates, the EU GDPR \cite{gdpr} proved to be  the most debated ground for legal scholars to argue about the enactment of a possible "\textit{right to an explanation}" within users affected by outputs of automated-decision making - and thus AI - systems \cite{wachterWhyRightExplanation2017, ebersExplainableAIEuropean2022}. In a similar vein, attention is now geared towards other EU regulations, e.g., the AI Act draft and how AI interpretability is thereby defined \cite{hackerVarietiesAIExplanations2022, sovranoMetricsExplainabilityEuropean2022, ebersExplainableAIEuropean2022}. 
Yet, up to now, to the best of our knowledge this attention has not addressed the AI regulatory trajectory within explainability informing policy documents. Similarly, no previous work has focused on comparing international approaches to AI explainability policies, reasoning over gaps and operationalization requirements. This perspective would offer an unprecedented background to visualize and evaluate regulatory approaches to AI explainability. This is the perspective we develop in the rest of this paper.



\section{Methodology: a thematic and gap analysis of AI explainability policies}
\label{sec:3}

To conduct our research, we first source policy documents from official governmental and affiliated agencies' websites of the  European Union (EU), United States (US), and United Kingdom (UK)\footnote{For the European Union, we refer to official websites such as e.g., the digital strategy of the European Commission \url{https://digital-strategy.ec.europa.eu/}, the official online law database Eur Lex \url{https://eur-lex.europa.eu/}, and standard bodies such as CEN \url{http://www.cen.eu/}, CENELEC \url{http://www.cenelec.eu/}, and ETSI \url{http://www.etsi.org/}. For the UK, we refer to \url{https://www.gov.uk/}, and related executive public bodies such as ICO \url{https://ico.org.uk/}, or research national institute as The Alan Turing Institute \url{https://www.turing.ac.uk/}. For the US, we refer to \url{https://www.whitehouse.gov/}, and governamental commissions such NSCAI \url{https://www.nscai.gov/}, or non-regulatory agencies such as NIST \url{https://www.nist.gov/}.}.
We collect four types of documents: 
{\textit{Communications}: related to AI governance strategies through public statements and releases;
\textit{Reports}: comprehensive studies, surveys, or official research papers that provide in-depth research information;
\textit{Regulations}: legally binding rules and guidelines that dictate how organizations must behave;
\textit{Standards}: technical specifications detailing implementation for AI explainability policies.}
We select the documents to review based on their level of relevance and availability of the data (e.g., document content under drafting might not be  disclosed to the public, but titles and expected releases might be). 
We  consider documentation produced from 2018 onwards to ensure that policies are tackling current AI developments. For the same reason, we consider the most up-to-date versions of the documents.
We exclude contexts where explainability or interpretability are presented outside of direct AI system involvement, e.g., when a regulation uses the terms in auditing procedures, thereby intending explanations as being required as justifications between humans over business conduct, etc. 
Based on the documents (refer to Appendix \autoref{tab:docs}, \autoref{tab:standards} for the complete list of documents reviewed), we  examine the regulatory landscape in the EU, US, and UK, exploring the distinct ways in which each jurisdiction has been addressing the regulation of AI over time, with a specific focus on explainability (\autoref{sec:4}). \textcolor{black}{Our thematic analysis identifies common themes across the collected documents encompassing not only "explainability" but also associated concepts such as "transparency" and "trustworthiness", while reporting similarities and differences in how they are accounted for. 
}
The last stage of our research is to conduct a gap analysis. While comparing themes across documents 
already reveals a number of gaps, we complement this understanding of the policies with research publications. Based on the themes above, we search for relevant  literature stemming from various research communities (algorithmic, human-computer interaction, ethics
), to identify misalignment with policies, calling for future work.

\section{Mapping the AI Regulatory Landscape for Explainability}
\label{sec:4}

\subsection{European Union: a risk-based approach to explainability for AI oversight}
\label{sec:4.1EU}

The European Union (EU) is driven by a commitment to strengthened legal frameworks for the development, deployment, and utilization of AI in line with its values, including fundamental rights, encoded in the trustworthiness and safety of AI systems \cite{european2021fosterinapproachtoai}. In this AI policy trajectory, the EU adopts a principle of regulatory proportionality, intended to be directly proportional to the severity of the consequences of the adverse effects produced by an AI system.

\subsubsection{Setting up AI strategies in the Data Economy (2018-2020)}
\label{sec:4.1.1}
In 2018, the European Commission (EC) acknowledged the potential of AI to drive economic growth and competitiveness through the \textit{Communication on Artificial Intelligence for Europe} \cite{CommunicationArtificialIntelligence2018}, outlining the first European strategy for its development and deployment. The document details a comprehensive approach to AI governance that balances promoting innovation with protecting citizens' rights and safety. The Communication states that the EU's approach to AI regulation is based on the principles of transparency, accountability, and human oversight.
Regarding transparency, the document approaches explainability as a high-level AI desiderata. The theme is mentioned under the principle of trust, increasing transparency, and mitigating bias and other risks, in a wider ethical and legal framework for AI development.

The first major discussion on the legitimacy of explanations over algorithmic decision-making systems, including AI ones, was found in the presumption of a ``\textit{right to an explanation}'' within the EU \textit{General Data Protection Regulation} (GDPR) \cite{gdpr} that came in force in 2018. The GDPR includes provisions for data subjects' rights such as accessing personal data and having them rectified in relation to automated decision-making. 
This is in conjunction to Article 22 and Recital 71, which state that individuals have the right not to be subject to a decision based solely on automated processing, including profiling, if it produces legal or similarly significantly effects on them. Art. 15(1)(h) emphasizes the importance of ensuring that individuals have \textit{``meaningful''} information about the logic involved in the decision-making process alongside \textit{``envisaged consequences''} of such processing for the individual. The provisions have been subject of an intense debate among experts, with some arguing that it provides a framework for ensuring transparency and explainability of automated decision-making \cite{malgieriAutomatedDecisionmakingEU2019, brkanLegalTechnicalFeasibility2020}, while others have criticized it for being too vague and difficult to implement in practice \cite{wachterWhyRightExplanation2017}. To notice, the phrasing of articles reduces severely the casuistry of its enforcement, not setting a baseline over typology and sufficiency of explanations, thus leaning towards an illusion of remedy rather than a burden of proof for legal recourse \cite{edwardsveale2018, ebersExplainableAIEuropean2022}.

As a final note, in 2020 the Commission published the \textit{White Paper on Artificial Intelligence} \cite{european2020white}. The document yet represents the first structural communication for an European strategy in AI to foster R\&D competitiveness. Even if explainability was barely mentioned once for deep learning (DL) outcomes, the Paper outlines objectives and policy options to build a digital \textit{ecosystem of excellence} through EU member coordination for research and innovation in AI across public sector, academia, industry, and small and medium enterprises (SMEs).

\subsubsection{Principles for AI trustworthiness: the HLEG guidelines (2019)}

The High-Level Expert Group on Artificial Intelligence (HLEG) established in 2018, published the following year their \textit{Ethics Guidelines for Trustworthy AI} \cite{EthicsGuidelinesTrustworthy2018}, the first organic attempt by the EC to define AI explainability.
Intended in a wider sociotechnical context, the term adopted was \textit{``explicability''}. It was defined as a principle connected to transparency, crucial to creating and maintaining users' trust in the AI system. The need for transparency corresponds to the epistemic condition to comprehend and challenge decisions of AI systems along three coordinates: \textit{Capabilities} - how system architecture is composed and what its functions are; \textit{Purpose} - to what purposes these capabilities correspond, established a priori; \textit{Decisions} - how and why are outputs processed. 
Worth noticing, a distinction is made between explainability per se and technical one. 
While the latter relates to system decisions to be understandable and traceable, the former includes also comprehending the purpose of the artifact in relation to the design choices followed. The context of an explanation is emphasized based on the expertise of the stakeholders involved directly (e.g., layperson, regulator, researcher). 
If capabilities and decisions are not deterministic, then indirect measures (e.g. stochastic model with surrogates) 
might assess impact and compliance with fundamental rights, measuring the need for explicability through a risk scale as informed by EU regulatory proportionality.


\subsubsection{Moving into AI regulations: the AI Act Draft and the AI Liability Directive (2021-2024)} 
\label{sec:4.1.3}
The path inaugurated by the HLEG and the White Paper informed EC's committees and their intense work on reinforcing the EU AI Act draft \cite{aiact_original} proposed in April 2021. The Act adopts a proportional risk-based approach to regulating AI systems, dividing them into three categories of risk with corresponding limitations, transparency requirements, and oversight mechanisms. Alongside definitions of risks assessments, transparency, and human oversight, the approach develops a taxonomy of AI actors within data quality requirements and technical documentation. The original draft included provisions for interpretability, specifically Article 13, requiring a sufficient degree of transparency mechanisms for \textit{high-risk} systems and the attachment of instructions for use containing relevant, accessible, and understandable information to users regarding the characteristics, capabilities, and limitations of performance of these systems. 
The Act has undergone several revisions as a result of the EC work, prior to inter-institutional negotiations being held in 2023. In April 2022, the IMCO-LIBE committee proposed initial changes to the regulations\footnote{Specifically within IMCO-LIBE, Amendment 43 to Recital 80(d) strengthened the EC's ability to access and understand databases, algorithms, and source codes. Similarly, Amendment 26 added Article 68(f), outlining the EC's ability to access a provider's premises and request explanations for the use of AI systems in analyzing documents and records. Amendment 265 to the creation of Article 68(g) established explainability as a mechanism for illustration and correction during the preliminary finding phase in cases of non-compliance.} surrounding AI explainability during inspections \cite{aiact_IMCO-LIBE}. In September 2022, the JURI committee stressed significantly explainability mentions with an Opinion \cite{aiact_JURI} addressing risk mitigation, user interaction and rights\footnote{For JURI, Recital 47(a) introduced transparency and explanation as a countermeasure for deterrent effects, Recital 80(a) expanded the possibility for an individual to invoke a right to explanation inspired by the EU GDPR, and Article 4(a) emphasized the value of transparency in AI system development in relation to traceability and explainability. Article 4(b) proposed to strengthen AI literacy strategies in organizations, while Recital 80(a) and Amendment 88 to Article 52(1) called for explicit communication of an AI system's capabilities and limitations. Additionally, Article 52 introduced the option of judicial redress for the harms and outputs of AI systems, including the \textit{right to an explanation}. Article 69 also provided for the explicit inclusion of a right to an explanation, specifying the decision-making procedure, main parameters of the decision, and related input data. Furthermore, Article 13 was strengthened with amendments 47-60 to increase understanding of how the system works for providers and end-users, as well as the data processed, in order to have a comprehensive view of how decisions affect them.}. 
Yet, those revisions were not substantiated in the final draft as presented by the Permanent Representatives Committee on November 25, 2022 \cite{aiact_generalapproach}, including several revisions aimed at ensuring the traceability and interpretability of high-risk AI systems. The changes include the introduction of indirect measures such as strengthened technical documentation requirements and instructions for use with illustrative examples, as well as guidelines for collecting and interpreting system logs. Article 13, originally stipulating a sufficient and appropriate degree of transparency for high-risk AI systems, was altered to include specific instructions for use and metrics related to the behavior of the system on certain groups of people and in relation to the sociotechnical context of adoption. Article 14(4)(c) was removed of the provision to have end users interpreting the characteristics of a system, 
 introducing it not as a requirement but as a possibility of consulting them through interpretation methods. Overall, final amendments 
weakened concepts of transparency and explainability, 
shifting the focus on ensuring the traceability and oversight of high-risk AI systems rather than empowering end-users' explanation demands.

As a final mention, in October 2022, the EC advanced a \textit{Proposal for an AI liability directive} \cite{ai_liability_direct} to define accountability allocation within non-contractual civil liability for damage involving AI systems. 
Whenever an allegedly damaged claimant suspects non-compliance of an AI system output, then it is required to establish a causal link as burden of proof. Art. 4(2)(b) details that if an AI system
is considered as high-risk, opaque, and complex (i.e., not allowing transparency requirements of the AI Act's Art.13), therefore explainability is mandated from an EU court not within the system (e.g., through XAI methods) but to the AI deployer through an order to disclose proportional evidence necessary (e.g., logs, documentation, and datasets). This is done to preserve confidentiality of trade secrets, as in Recital 16, it is said that the AI Act does not specify any right for an injured person to access that information.

\subsection{United States: explainability research game changer, loose policy measures}
\label{sec:4.2USA}
In the United States of America (US), the approach to AI regulation has relied mostly on self-regulation, where industry stakeholders develop best practices and guidelines for the use of AI \cite{partnership} even if critics pointed a loose legal oversight \cite{CAIDPIndex2022}.
The US approach to AI explainability had a pioneering role through the announcement in 2016 of the Defense Advanced Research Projects Agency (DARPA) of the federal funding BAA 16-53, inaugurating the first research program in \textit{Explainable AI} (XAI) \cite{XAIdarpa}.

\subsubsection{The road to an AI leadership}
Strong in its position of global technological powerhouse, the US AI regulatory trajectory started to gain momentum in February of 2019, when the Executive Order (EO 13859) on \textit{Maintaining American Leadership in AI} was announced by the White House \cite{MaintainingAmericanLeadershipinArtificialIntelligence}. Despite the nature of order addressing specifically a national R\&D strategy to keep up its global competitiveness, the policy tone thereby advocated for an ethical, responsible, and transparent development of AI. 
This balance informed the \textit{Guidance for Regulation of AI Applications} issued by the Office of Management and Budget (OMB) later in 2020 \cite{GuidanceforRegulationofAIapplications}.
The Guidance stated that AI systems should be designed accordingly to be trustworthy, auditable, and thus explainable - especially for the mentions to transparency and disclosure reported at section (8.). Interpretability of AI systems is also briefly announced in the Appendix  to enhance transparency for oversight. But rather than leveraging XAI methodologies, indirect regulatory processes are proposed for that aim, i.e., impact analysis, public consultation, and risk assessments.
A more structured answer to EO 13859 was elaborated by NIST. Indeed, the Order mandated the agency to develop a plan \cite{NISTFedEng} to tackle Federal priorities for a robust and safe AI R\&D while promoting international standardization activities. In the plan, explainability is ascertained below the concept of trustworthiness, one of nine key areas of focus identified for AI standards. 
These governmental policy communications found a wider output in the \textit{Final Report} \cite{NSCAI2021FinalReport} issued by the National Security Commission on AI (NSCAI) March 1st, 2021. The federal commission, created in August 2018 and dismantled in October 2021, had the mandate to advice the US President and Congress on AI R\&D for national security and defense needs, reflecting the willingness to maintain its technological competitiveness through workforce development, international cooperation, and AI ethics. 
There, 
explainability is only mentioned among areas of R\&D, alongside 
transparency and accountability of AI deployment in national security applications. 
These strategic documents reflect a lack of informed perspective over AI explainability for civil rights, filled only partially by the  release in October 2022 of the \textit{Blueprint for the AI Bill of Rights} by the Office of Science and Technology Policy (OSTP) \cite{blueprintforanaibillofrights}. Rather than  industry oriented, the Blueprint can be seen as a civil rights framework for ensuring that automated decision making systems (ADMs) are used in ways that respect American values such as privacy, autonomy, and other civil liberties. Under the principle of \textit{Notice and Explanation}, automated decisions shall be justified through "\textit{clear, timely, understandable, and accessible}" use in connection to valid explanations tailored to purpose, audience target, and level of risks.

In terms of bills, the US R\&D AI trajectory found legislative output in the enactment of the \textit{National AI Initiative Act} on January 1st,  2021 \cite{USAnationalAIiniatitiveAct}. The Act remarked the duty (Sec. 22A(c)(2)) to establish, within NIST a voluntary risk management framework by 2023 \cite{NIST-AIRMF1.0}, detailing also common definitions and characterizations of aspects of AI trustworthiness such as explainability.
Yet for civil rights and litigation in March of 2022 a relevant bill was introduced for possible enforcement of explanations within ADM/AI systems. In Section 4 of the \textit{Algorithmic Accountability Act} \cite{USAAlgoAccountAct2022}, companies deploying such systems will be required to perform impact assessments. Among the most impacting provisions figure requirements for business explanations over data collection and maintenance (Sec. 4(7)(A)(ii)) and evaluation standards. Also, there is need to deliver end-users explanations over system features contributing to the decision output, as well as overall information about the system and process (Sec. 4(8)(B)(i)). Interestingly, provision Sec. 4(11)(C) further addresses the need to engage stakeholders to provide feedback on improvements for the ADM/AI system also for their explainability. To conclude, in context of Commission oversight, Sec.5(1)(H)(i) mandates to submit documentation from the impact assessment over system transparency and explainability measures.

\subsubsection{Explainability within NIST}
\label{sec:4.2.2}
Despite the loose mentions to explainability in AI R\&D as shown in policy communications, the theme is considered part of NIST's "\textit{Fundamental AI Research}" \cite{NISTxaiWG}.
The working group for explainability has so far released two publications in 2021. 
In April, a first paper (NISTIR 8312)\cite{NIST-8312} reviewed human psychological traits that characterize users within algorithmic explanations. The document, unique in its genre, expose over psychological properties of interpretation and explanation related to human mental representations. The former term is said to be useful for policymakers and general users for AI system oversight, while the latter is valuable to developers for debugging and design improvement. Yet, distinction is loose, and the value of individual differences is underlined, as well as the recommendation to design interpretable algorithms contextualizing data in relation to human cognition representations.
This perspective in cognitive psychology and user interaction informed the second paper (NISTIR 8367) \cite{NIST-8367} in September 2021 on four principles for explainable AI systems. Through the acknowledgment that explanation types should differ based on users, the principles indicate AI explanations to be designed as meaningful to humans, accurate over system inherent process, and expressing system capabilities and knowledge limits.

As set in the \textit{National AI Initiative Act}, these reports informed the release in January 2023 of the first version of the \textit{ Risk Management} for AI systems (AI RMF 1.0) \cite{NIST-AIRMF1.0}. The framework
incorporates trustworthiness considerations into the design, development, use, and evaluation of AI products, services, and system. Explainability and interpretability are named as characteristics of \textit{AI Trustworthiness} in Section 3. These characteristics are evaluated within a few AI risks and trade-offs, e.g., enhancing system interpretability against predictive accuracy or achieving privacy. Yet, the framework just provides a coarse overview of the terminology, referencing the previous NIST's papers, without detailed descriptions of explainability methods nor specific requirements for users and contexts of AI systems, thus lacking a proper informed baseline. Yet, in the second part of the framework, four functions are advanced to evaluate and mitigate AI risks (i.e., \textit{Govern}, \textit{Map}, \textit{Measure}, \textit{Manage}). Explainability is found below \textit{Measure 2.9} function for quantifying AI risks. In other words, the concept diverges from previous NIST's approaches being narrowed down in its functionalities, as related only to the evaluation phase for trustworthy characteristics. Similarly, interpretation is briefly considered for AI system output within its context, without specifying further baseline or approaches also directed to different targets, e.g.,  system capabilities and complexity.

\subsection{United Kingdom: explainability guidance for government and industry}
\label{sec:4.3UK}
The approach of the United Kingdom (UK) to AI regulation is still evolving, and ongoing discussions are taking place on the need for specific legislation to address the risks and challenges posed by AI, with a focus on explainability addressed foremost to industry adoption. Yet, the UK government is well grounded ensuring that AI systems are developed and deployed in a way that is safe, trustworthy, and respects the rights and interests of individual \cite{UkNAtionalAIStrategyActionPlan2022}.

\paragraph{Guiding organizations for explanability}
During the implementation period of Brexit, in 2018 the UK government established the Centre for Data Ethics and Innovation (CDEI), the first governmental body worldwide to advise on the ethical and societal implications of AI and data-driven technologies. In 2019, the UK released a \textit{National Data Strategy} \cite{UkNatDataStrategy2019} advocating for the adoption and use of safe and explainable data sources, in parallel to an AI Sector Deal \cite{UkAISectorDeal2019}. The Deal outlined a public AI R\&D strategy to enhance its market competitiveness while reinforcing digital infrastructures. In the plan, it was announced the intention to set up a research collaboration between the Alan Turing Institute (ATI) and the Information Commissioners Office (ICO) for a guidance in explaining AI decisions. In 2020, that guidance was released in a well structured report named \textit{Explaining Decisions with AI} \cite{ExplainingDecisionsMade2020}. In that particularly prolific year for the ICO, it also released reports on AI auditing frameworks \cite{ICOGuidanceontheAIauditingframe2020} and AI data protection \cite{ICOGuidanceontheAIanddataprotection2020}. The guidance on explainability is the first comprehensive governmental technical report ever produced detailing AI explainability definition, methods, process, and impacts. The document is structured in three parts, respectively addressed to compliance teams, technical teams, and management. This reflects the priority given by ICO \& ATI to provide industry guidance on explainable AI. Yet, their major contribution is to be found in the establishment in 2019 of \textit{Project ExplAIn} \cite{projectexplain}, the first engagement strategy on AI explainability for organizations, implementing the guidance through workbooks and workshops.


\paragraph{Leading AI industry standards}
The UK is also taking a novel approach to widening participation in AI standards, led by the Alan Turing Institute, the National Physics Laboratory (NPL), and the British Standards Institute (BSI). To notice, in May 2021 a guidance on ethics, transparency, and accountability of ADM systems was released by the Central Digital \& Data Office (CDDO), the Cabinet Office, and the Office of Artificial Intelligence \cite{UK2021EthicsTransparency}. Their major output is the development of a cross-government standard hub for algorithmic transparency for government departments and public sector bodies (Pillar 3 of the National Strategy plan) \cite{UkNAtionalAIStrategyActionPlan2022} announced in November 2021 \cite{UK2021algotransaprencyrecordingstandard}. Informed by its close collaborations within AI standardization bodies in the EU and worldwide, the UK plans to integrate future standards in its AI innovation strategies \cite{UkNAtionalAIStrategyActionPlan2022, UkEstablishingaproinnovationregAI2022}. This direction found output in the first algorithmic transparency standard for government departments and public sectors in November 2021 \cite{UK2021algotransaprencyrecordingstandard} alongside a report from The Alan Turing Institute \cite{AlanTuringCommonRegCapAI} in July 2022, calling for a coordinated approach to increase AI readiness and proposing the creation of an \textit{AI and Regulation Common Capacity Hub} to facilitate regulatory collaborations for policymakers. 

\subsection{A perspective on standards for explainability}
\label{sec:4.4standard}
The EU has a complex system of actors involved in the development and implementation of standards for AI regulation. The EC serves as the executive branch and plays a crucial role in shaping the regulatory framework. The European Standards Organizations (ESOs) such as CEN, CENELEC, and ETSI are responsible for creating standards in support of the EU AI Act (AIA)\footnote{In the context of AI regulation, the CEN (\textit{European Committee for Standardization}) and the CENELEC (\textit{European Committee for Electrotechnical Standardization}) play a crucial role in defining the technical standards and safeguards needed for the effective implementation of regulations. CEN is responsible for creating European standards in the areas of consumer goods, engineering, healthcare, environment, and other related fields, while CENELEC is responsible for setting standards in the areas of electrotechnical engineering and applications. Under the New Legislative Framework in the EU, standards can be referenced in the Official Journal of the European Union, in order to provide presumption of conformity with particular legislation. These standards provide the technical details to support conformity assessment of products and services. In terms of digital society, CEN-CENELEC often expand their work in the international landscape, collaborating with ISO and IEC.} alongside a variety of stakeholders including national standards bodies, European stakeholder organizations, and harmonized standards consultants as outlined in a report from 2021 from the EC on the EU AI standardization landscape \cite{EUJRCAIWatchStandLadnscape2021} and in the 2022 \textit{Rolling Plan for ICT Standardization} \cite{europeanrollingplanforictstandard2022}.
The EC, on December 2022, provided a draft standardization request to CEN-CENELEC outlining requirements for standards to support presumption of conformity \cite{ecdraftstandardisationrequest}.  
CEN-CENELEC may choose to develop their own standards, but can also adopt the ISO/IEC standards as sufficient to meet the needs of the standardization request. For further development within AI explainability, attention should be given to CEN-CENELEC's Joint Technical Committee 21 \textit{'Artificial Intelligence'} (JTC 21) and the AI work program communicated in 2020, where explainability was mentioned among key research themes for standards development \cite{cencenelecroadmap}.
With CEN-CENELEC, the European Telecommunications Standards Institute (ETSI) is the third body that will work on advising for the AI Act. Since September 2019, the ETSI has a Strategic Advisory Board on Artificial Intelligence (SAI) \cite{ETSIGRSAI}. The ETSI ISG SAI has identified several areas for standardization, including transparency and explainability, and aims to produce several deliverables related to the AI Act draft \cite{etsiactivitiesaia}, including an existing deliverable on the \textit{Experiential Networked Intelligence} (ENI) system architecture and two forthcoming deliverables on explainability and transparency of AI processing [ETSI ISG SAI GR 007] \cite{etsi007} and traceability of AI models [ETSI ISG SAI GR 010]\footnote{The work on these deliverables is available within the ETSI Technical Committee on Intelligent Transport Systems.}. 

On the US side, NIST has produced two papers to reflect over principles and methods for explainability informed by research perspectives \cite{NIST-8312, NIST-8367}, yet this seem not to have been transposed consistently in the \textit{Risk Management Framework v1.0} \cite{NIST-AIRMF1.0}. Despite the non-binding nature of NIST, their guidance could be in the near future reinforced by standardization initiatives from the US Federal Trade Commission (FTC) given their activities addressing irregular ADM practices, and by future work for enacting the Algorithmic Accountability Act \cite{FTCtrade, protocolalgo}.  

Similarly to the US, the UK have not substantiated yet in binding standards or regulations affecting AI explainability. 
Also informed by the recent settlement with the Standard Hub \cite{UK2021algotransaprencyrecordingstandard, AlanTuringCommonRegCapAI}, the UK is able to take an equivalent approach to the EU in selecting standards that provide presumption of conformity. Despite leaving the European Union, UK remains member of ESOs such as CEN/CENELEC\footnote{By promoting AI standardization's, the UK is also involving organizations that do not normally have the time or resources to contribute to the development process. In January 2023, the Standards Hub led a full day workshop attended by 20 selected experts from government and industry. Experts from ISO/IEC SC 42 and CEN/CENELEC attended to lead a discussion on the definitions of explainability, interpretability, and transparency. This resulted in a written contribution that is being submitted to SC 42 for future consideration.} while being actively engaged in the draft of international AI standards \cite{UkEstablishingaproinnovationregAI2022}.

Yet, as mentioned earlier, governments tend to refer to international standardization activities in AI provided by the International Standard Association (ISO) jointly with International Electrotechnical Commission (IEC), alongside the Institute of Electrical and Electronics Engineers (IEEE).
The technical committee ISO/IEC JTC 1/SC 42 '\textit{Artificial Intelligence}' \cite{ISOIECJTC1SC42} focuses on the standardization AI program within ISO/IEC Joint Technical Committee JTC 1. As the central advocate for AI standardization, this committee provides direction to JTC 1, IEC, and ISO committees working on the development of AI applications. 
Connected to the principles of trust and transparency, their document ISO/IEC TR 24028:2020 \cite{ISOIECTRR240282020} published in May 2020 considers explainability as a mitigation measure to AI vulnerabilities and threats, surveying existing approaches in explainability methods and evaluations. Currently under development and expected to be released during 2024, document ISO/IEC AWI TS6254 \cite{ISOIECAWITS6254} will describe explainability methods for 
AI systems accordingly to different stakeholders (i.e., academia, industry, policy makers, end-users among others), while ISO/IEC AWI 12792 \cite{ISOIECAWI12792} will define a taxonomy of AI system transparency information elements.
For IEEE, the standardization activities are more scattered and not piloted by a single committee. For such a reason, different documents affecting AI explainability are proposed, such as a standard for algorithmic bias consideration over output interpretability (CC/S2ESC/ALGB-WG P7003) \cite{IEEEP7003}, a guide for XAI techniques, application, and evaluation (C/AISC/XAI P2894) \cite{IEEEP2894}, a standard for transparency of autonomous systems (VT/ITS 7001-2021) \cite{IEEE70012021}, and a standard for mandatory requirements to recognize an AI system as explainable (CIS/SC/XAI WG) \cite{IEEEP2976}. Interestingly, the XAI guide was supposed to be released in 2022 but no further  updates are presented, while the standard is expected for July 2024 at the latest. 



\section{Discussion: regulating explainability}
\label{sec:5}
The European Union has made significant efforts in policy communication and legislation related to data and AI, but the GDPR and the proposed AI Act do not contain clear requirements for interpretability and explainability for end users. There is some emphasis on explainability for oversight purposes, but the lack of technical reports and standards in this area highlights a need for further development. In the United States, there is significant investment in AI R\&D and some mention of explainability in relation to civil rights, but the current regulatory landscape is largely focused on driving innovation rather than protecting human rights. The Algorithmic Accountability Act draft includes some provisions for end-user rights to explanation, but the policy focus remains on innovation. In the United Kingdom, the regulatory landscape for AI is limited to data protection, and there is a lack of clear enforcement or standardization in the area of explainability. Currently, the focus seems to be on providing guidance to industry and promoting innovation, with limited attention paid to end-user rights and protections. These focuses call for a more balanced approach that balances commercial interests with human rights and the need for trustworthy AI.

\subsection{Main themes stemming from our analysis}\label{subsec:themes}


Governmental approaches to AI explainability have been marked by a growing recognition of the importance of interpretability and transparency in AI systems, and a commitment to the development of explainable AI systems that can be audited and held accountable in the wider governance framework of AI risks management. NIST or ICO approaches to explainability research and organizational implementation yet did not substantiate in actionable regulatory standards. 
Yet, provisions such as the EU GDPR, \textit{AI Act Act} draft,\textit{ AI Liability Directive} proposal, and also the US\textit{ Accountability Act} draft might be considered a first legal baseline for explainability operationalization, but technical requirements might benefit from less coarse specifications spanning also towards end-user services, and not just oversight. 
Indeed, 
as witnessed by casuistry and feasibility within the EU GDPR 
(Sec. \hyperref[sec:4.1.1]{4.1.1}), 
challenges might arise in establishing a legal baseline for explanations, balancing between explainability method criteria, model inherent complexity, stakeholders interests and expertise, as well as contextual organizational factors that might introduce further tensions connected to enhanced transparency. 
%
These organizational factors, such as trade secret integrity, intellectual property rights, and privacy of third parties could be intended as safeguards and thus incentivize organizations' R\&D, without feeling pressured to be continuously subject to explanation requirements. In this vein, an organization might feel less threatened by not being mandated for explanations over their AI system process and output to users, since the latter might deploy them as a burden-of-proof under litigation \cite{bordtPostHocExplanationsFail2022}. Rather, they might prefer being subject to clear regulatory practices to indirectly assess that AI models, despite their inherent complexity, can be considered safe, fair, and thus compliant with the rule of law.

Even before that, bills seem to leave 
open the assessment of transparency and context of use for explanations to end users, upon provider discretion e.g., in the EU AI Act, at the discretion of the provider of high-risk systems only 
(Sec. \hyperref[sec:4.1.3]{4.1.3}).
Such ambiguity can stem from a wider sense of terminology ambiguity that characterized the production of communications and reports across governments and commissioned standardization bodies. As an example, we noted that NIST's papers on explainability provided an in-depth research analysis, but this seemed not to substantiate in their \textit{Risk Management Framework} nor the White House's \textit{Blueprint} (Sec. \hyperref[sec:4.2.2]{4.2.2}).
On this line, we could further trace back this ambiguity to the lack of official coordination among agencies and their inner working groups, as for example within IEEE where two different working groups have been proposed to respectively draft a guide (C/AISC/XAI with P2894 \cite{IEEEP2894}) and a standard (CIS/SC/XAI WG with P2976 \cite{IEEEP2976}) to AI explainability (Sec. \hyperref[sec:4.4standard]{4.4}).
In the rest of the discussion, we further detail tensions and limits we detected in policies and standards as informed by academic research. 

\subsection{Accounting for the recent nascence of the concept of explainability in research}

One important limitation of current documents is that, while they recognize the importance of explainability for civil right enhancement, their high-level mentions in policy communications hint to the lack of an informed perspective of explainability as a research object still nascent, novel, and complex, and far from being a "solved problem". As is currently, an AI developer cannot easily implement "explainability" in their systems, as the meaning of it and the methods for it might not exist. If a body envisions to develop standards for explainability, once again, the task could not be performed yet entirely for the reasons and tensions we envision below.

\paragraph{Theoretical understanding of explainability}
Explainability being a nascent concept, it remains unclear what should be considered a \emph{good} explanation \cite{lage2019evaluation,hase2020evaluating,dhanorkar2021needs}. Early HCI works have shown that explainability is contextual, different stakeholders might need different types of explanations depending on their purpose \cite{liao2022connecting,ehsan2021explainable},  on the domain of application, as well as on a plethora of human factors (e.g., AI literacy, cultural background) that still need to be characterized \cite{balayn2022can}. On the other side, policy documents often mention explainability in AI strategies without providing any specification in terms of stakeholders, purpose, nature of explanations, etc. Leaving these concepts undefined in policies while they are also yet to be comprehensively understood from research, allows for ambiguity that might be beneficial to the development of new methods, but also detrimental to their responsible use.

\paragraph{Practical feasibility of implementing explainability}
Practically, the designers and developers of an AI system might encounter obstacles when building a system with explainability in mind. 
While policy documents discuss three types of explainability for an AI system, the vast majority of research papers has only focused on one of them, the technical explainability. Besides, explainability methods are data-, task-, and algorithm- specific. Not all currently used AI systems in production have now been associated with explainability methods, as a large majority of the research now focuses on DL technology (whereas organizations still rely greatly on traditional ML models).
Hence,  one would not necessarily be supported in developing the right types of explainability.
This narrow focus probably stems from the way the research community is organized, prioritizing algorithmic research especially around DL, to other types of research in terms of rewards and venues for publication, methodologies employed and taught, recognition from peers, and organizational incentives \cite{sambasivan2021everyone}.
Furthermore, for the explainability methods that do exist, it is now well-known that they suffer from several issues hindering their use in practice. They are often of low-fidelity \cite{adebayo2018sanity}, brittle to various types of perturbations such as adversarial attacks \cite{slack2020fooling,ghorbani2019interpretation,bordtPostHocExplanationsFail2022}, and inconsistent \cite{lee2019developing}, not always allowing to observe all issues (e.g., spurious correlations) a ML model might suffer from \cite{adebayo2022post}. Yet, it remains challenging for researchers to develop better explanations, as developing appropriate objectives and benchmarks for explainability is not even a solved problem until now \cite{mohseni2021quantitative}.

\paragraph{Usability of explainability for stakeholders}
Assuming that there would exist methods for making an AI system explainable, additional challenges arise. 
On one side, AI designers and developers might not be aware of these methods (the research / practice disconnect is a well-known one, more broadly than simply for AI) and might not be able to correctly use them for their system (methods might require more or less algorithmic knowledge, coding abilities, etc.) \cite{belle2021principles,bhatt2020explainable,balayn2022can}. 
On the other side, it is also well-studied that those who exploit the explanations resulting from the explainability methods implemented within the AI system might not be well-supported to do so \cite{bhatt2020explainable,krishna2022disagreement}. Again, they might lack knowledge to understand them \cite{jesus2021can}, and might fall into traps from various cognitive biases \cite{mohseni2021quantitative,hase2020evaluating,krishna2022disagreement} leading them to blindly trust the AI systems. 
For instance, it has been found that explanation receivers might be deceived with tailored explanations  sounding factual from an epistemic perspective  and satisfying prior beliefs, leveraging conditions of confirmation or automation bias \cite{balagopalan2022road} or illusions of explanatory depth \cite{chromikexplanatorydepth}.
This might turn out to be a costly organization choice if a biased, unfactual explanation could be deployed as burden of proof during litigation by an end-user, e.g., as reported in the EU AI Liability Directive.


\subsection{Accountability allocation of explanations}
Besides the theoretical and practical challenges discussed above to develop and use explainable AI systems, another set of concerns revolve around organizational obstacles towards explainable AI. Various tensions exist with making a system explainable. 
As also reported by NIST, at an algorithm-level it is now well understood that a complex, intricate, contextual, trade-off exists between explainability and other desirable properties of an AI system such as accuracy \cite{bell2022s}, or privacy-preservation \cite{budig2020trade}. As it is often said that organizations might prefer accuracy as it allows for more effective and efficient business processes, this might come as an obstacle towards making models explainable. 

\paragraph{Discretion of the service providers}
More broadly, several publications have posed the existence of organizational factors \cite{madaio2020co,rakova2021responsible} that might become obstacles to developers willing to make models more explainable, i.e., absence of incentives, scarcity of time, burdensome computational costs of explanations might be inefficient \cite{chenshapley}. The publications report these factors for fairness objectives, but these could be easily transposed.
On the other hand, policy documents (e.g., Art. 13(1), EU AI Act) 
appear to leave at
the discretion of the service providers to design AI systems for explainability, and especially for high-fidelity, high-explainaibility-level, and usable explanations. Similarly to what is currently discussed around the regulations of AI towards fairness \cite{balayn2021beyond}, 
uninterested providers might 
implement the easiest solutions, that might not actually be enough for supporting explainability in effect.

\paragraph{Gaming opportunities and ethics washing}
Connected to implement the easiest solution to ensure legal compliance, this practice can be ascertained to the concept of \textit{ethics bluewashing} \cite{Floridi2019}. Given the discretion upon organizations reported in the analyzed policies, AI explanation typologies and targets could spotlight part of data and system design that can be considered as desirable and compliant. Explanations could be deployed to vaguely describe AI system properties rather than to inform on relevant actions the decision-subject shall take to remedy to her condition, or on what design rationale informed the development and maintenance of the AI system \cite{robbins2019misdirected, CABITZA2023118888}.
Similarly, \textit{ethics dumping} or \textit{shopping} of explanations could justify organization behaviour while appealing to good practices or laws present in a single region, being endorsed not by international recognized standardization bodies but other stakeholders with more "appealing" code of ethics and standards \cite{wagner2018ethics, Floridi2019}. 
Explanations should be then operationalized accordingly to the rule of law affecting recipients of explanations, establishing clear baselines to reprove accountability measures, e.g. in the US \textit{Accountability Act} and EU \textit{AI liability} drafts. Yet, obliging to legal baselines should be not considered equivalent to truly endorse "ethics best practices" \cite{bietti2020ethics, green2021contestation, hickok2021lessons}. Future enactments of AI regulations paired with data and service ones (e.g., EU will enforce its \textit{Digital Services Act} \cite{digital_services_act} in 2023) will constitute an interesting benchmark to evaluate ground-truths of algorithmic explanations, hopefully contributing to allocate responsibilities over non-factual information.

\section{Avenues for Future Work}
\label{sec:6}

\subsection{Methodological limitations of our research}

Given our analysis scope, we centered our focus only on the EU, US, and UK regulatory landscapes, thus not fully capturing the global landscape of AI explainability policies and recommendations
outside of those regions, e.g., G20 members and other countries with minor impact in drafting AI policies yet affected by them \cite{mohamed2020decolonial, png2022globalsouth}.
While the analyzed governmental and official standardization bodies play a crucial role in shaping the legal and technical landscape of AI explainability, future research should account the influence of further actors, such as industry associations, civil society organizations, and academic experts. 
In terms of influences, our analysis tackles explainability in the context of AI policies. Future work should account other important factors 
interrelating to that 
in practice, e.g., data and digital platform laws, trade secrets, and privacy concerns. 

We also note how, at this point in time, our analysis is based on a relevant quantity of documents under development. Therefore, final versions of the discussed drafts here might differ in the next future, presenting new amendments and unforeseen tensions within AI explainability policies. As the AI field is evolving, our analysis and recommendations should be re-evaluated as new regulations and policies will be mirrored accordingly.
Future research should tackle these issues by examining how they interact with explainability tensions and recommendations hereby outlined.

\subsection{Recommendations for future work}
\begin{itemize}[leftmargin=*]
\item \textit{Disambiguating terminologies.} Across policy documents, we note a great diversity of terminological ambiguities. Most documents do not clearly define the scope of technological systems they tackle (e.g., mentioning either ``AI'', ``ML'', ``ADM'', etc.), the relevant concepts surrounding explainability such as its definition and typology, nor its purpose in relation to its various stakeholders (e.g., ``end-users'' is often used but it is unclear  whether it refers to end-users of the AI system, or of the explanation which would encompass decision subjects, developers, assessors, etc.). Prior works have already illustrated terminological confusions in AI ethics and policy \cite{mulligan2019thing, krafftetaldefining}. We second their suggestions, and emphasize the need for precise concept definitions in policy documents to foster actionable recommendations.

\item \textit{Accounting for the complexity of explainability.} As discussed in \autoref{sec:5}, explainability is a recent concept, that is complex and prone to be gamed, and still iterated over in research. Hence, we recommend future policy documents to remain cognizant of these particular characteristics also discussed around AI fairness \cite{balayn2021beyond} and privacy \cite{gurses2011engineering,danezis2014privacy}, and to strive for flexibility to adapt to changes, while also remaining specific enough to enforce relevant constraints.

\item \textit{Fostering further research.} We emphasize the importance of a reciprocal influence between policy and research. While we identified limitations of current policy documents based on research outputs (and recognize the complexity of the problem since it requires collaborations between various roles, technical and legislative, researchers and policy-makers, etc.), we also propose that research should 
cater for prioritizing needs stemming from these documents. This might entail not developing more efficient AI systems but re-focus towards different types of explainability.

\item \textit{Looking ahead towards explanations for AI innovations.}
To conclude, a brief mention needs to be addressed for regulations  to be "future-proof" within new AI advanced systems. These systems, such as e.g., generative and foundation models, might pose exceptional challenges within their interpretability, being composed by an aggregation of multiple AI systems stacking up complexity of inspection \cite{foundationmodels}.
Similarly, explanations provided by large language models might not reflect factual information, delivering explanation for user recommendations without previous fact-checking \cite{bendershahsituatingsearch}. This might lead to cascade effects of societal harms not addressed by policymakers, connected to misinformation and public opinion deception, potential eroding adoption of AI explanations due to jeopardized public perception of their trustworthiness and stricter deployment \cite{Fenwick2018, smuha2021beyond}. Similarly, AI systems built via novel AutoML tools \cite{karmaker2021automl} might require different types of explanations that need to be accounted for in future policies.
\end{itemize}

\section{Conclusion}
\label{sec:7}

In this study, we rigorously surveyed policy documents and standards stemming from the EU, US, and UK, and contrasted them with research publications around AI explainability. We contribute the first, valuable, starting point for understanding the current state of AI explainability in policies and standard.
We particularly found that policy trajectories prioritize AI innovation under a risk management lens rather than truly empowering users with explanations. Despite the rising attention for civil rights, we find that documents might fall short in addressing complexity and sociotechnical impact of explainable AI, leading to missed-opportunities over its development and implementation.
With this awareness, we advocate a series of recommendations for future policies.
We hope that, in the long term, this informed perspective will foster new standards and guidelines specifically tailored to the different contexts in which explainable AI becomes relevant, and will inspire further necessary research directions. 

\begin{acks}
Funding contribution from the ITN project NL4XAI (\textit{Natural Language for Explainable AI}). This project has received funding from the European Union’s Horizon 2020 research and innovation programme under the Marie Skłodowska-Curie grant agreement No 860621. This document reflects the views of the author(s) and does not necessarily reflect the views or policy of the European Commission. The REA cannot be held responsible for any use that may be made of the information this document contains.
\end{acks}

\bibliographystyle{ACM-Reference-Format}
\bibliography{sample-base}

\newpage
\appendix
\section{Appendix - Table of Policy References Consulted}

\begin{center}
\begin{table}[h!]
  \caption{A non-exhaustive list of policy documents such as communications, reports, and regulations involving AI Explainability. 
  }
  \label{tab:docs}
  \begin{tabular}{p{0.6cm}p{4.2cm}p{4.2cm}p{4.2cm}}
    \hline
    \textbf{Area} 
    & \multicolumn{1}{c}{\textbf{Communications}} & \multicolumn{1}{c}{\textbf{Reports}} & \multicolumn{1}{c}{\textbf{Regulations}} \\
    \hline
    \addlinespace[1ex]
    \textbf{EU} 
    & 
    \begin{itemize}[label=\textbullet, leftmargin=*, labelsep=1mm]
        \item 2018, EC AI for Europe \cite{CommunicationArtificialIntelligence2018}
        \item 2020, White Paper on AI \cite{european2020white}
        \item 2021, EC Fostering a European approach to AI
        \cite{european2021fosterinapproachtoai}
        \item 2021, EC Coordinated Plan on AI 2021 Review
        \cite{european2021coordinatedplanonai2021review}
        \item 2022, EC Rolling Plan for ICT Standardisation - AI
        \cite{europeanrollingplanforictstandard2022}
    \end{itemize} 
    & 
    \begin{itemize}[label=\textbullet, leftmargin=*, labelsep=1mm]
        \item 2019, HLEG Guidelines for Trustworthy AI \cite{EthicsGuidelinesTrustworthy2018}
        \item 2020, HLEG Assessment List for Trustworthy Artificial Intelligence (ALTAI) \cite{HLEGALTAI2020}
   
        \item 2021, JRC AI Standardization Landscape \cite{EUJRCAIWatchStandLadnscape2021}
    \end{itemize}  
    &  
    \begin{itemize}[label=\textbullet, leftmargin=*, labelsep=1mm]
        \item 2018, EU GDPR [Enactment] \cite{gdpr}
        \item 2021, AI Act [First Draft] \cite{aiact_original}
        \item 2022, AI Liability Directive [Proposal] \cite{ai_liability_direct}
        \item 2022, AI Act [Final Draft - General Approach] \cite{aiact_generalapproach}
    \end{itemize}
    \\
    \cline{2-4}

    \textbf{US} 
    & 
    \begin{itemize}[label=\textbullet, leftmargin=*, labelsep=1mm]
        \item 2019, WH - Executive Order on Maintaining American Leadership in AI \cite{MaintainingAmericanLeadershipinArtificialIntelligence}
        \item 2020, WH - OMB -  Guidance for Regulation of AI Applications \cite{GuidanceforRegulationofAIapplications}
        \item 2022, Blueprint for AI Bill of Rights \cite{blueprintforanaibillofrights}
    \end{itemize} 
    &
    \begin{itemize}[label=\textbullet, leftmargin=*, labelsep=1mm]
        \item  2021, NSCAI Final Report \cite{NSCAI2021FinalReport}
        \item  2021, NIST Federal Engagement Plan \cite{NISTFedEng} 
        \item  2021, NISTIR 8367 \cite{NIST-8367}
        \item  2021, NISTIR 8312 \cite{NIST-8312}
        \item  2023, NIST AI RMF v1.0 \cite{NIST-AIRMF1.0}
    \end{itemize}  
    & 
    \begin{itemize}[label=\textbullet, leftmargin=*, labelsep=1mm]
        \item 2021, National AI Initiative Act [Enactment] \cite{USAnationalAIiniatitiveAct}
        \item 2022, Algorithmic Accountability Act [Draft]
        \cite{USAAlgoAccountAct2022}
    \end{itemize} 
    \\
    \cline{2-4}
  
    \textbf{UK} 
    & 
    \begin{itemize}[label=\textbullet, leftmargin=*, labelsep=1mm]
        \item 2019, National Data Strategy \cite{UkNatDataStrategy2019}
        \item 2019, AI Sector Deal \cite{UkAISectorDeal2019}
        \item 2019, ICO \& Alan Turing Institute Project ExplAIn \cite{projectexplain}
        \item 2022, National AI Strategy - AI Action Plan \cite{UkNAtionalAIStrategyActionPlan2022}
        \item 2022, Establishing a pro-innovation approach to regulating AI \cite{UkEstablishingaproinnovationregAI2022}
    \end{itemize} 
    &
    \begin{itemize}[label=\textbullet, leftmargin=*, labelsep=1mm]
        \item 2020, ICO, Guidance on the AI Auditing Framework \cite{ICOGuidanceontheAIauditingframe2020} 
        \item 2020, ICO, Guidance on AI and Data Protection \cite{ICOGuidanceontheAIanddataprotection2020}
        \item 2020, ICO \& Alan Turing Institute, Explaining Decisions with AI \cite{ICOExplainingdecisionsmadewithAI2020}
        \item 2022, Alan Turing Institute, Common Regulatory Capacity for AI \cite{AlanTuringCommonRegCapAI}
    \end{itemize}  
    & 
     \begin{itemize}[label=\textbullet, leftmargin=*, labelsep=1mm]
        \item 2018, Data Protection Act [Enactment] \cite{Ukdataprotectionact2020}
    \end{itemize} 
    \\
    \hline
  \end{tabular}
\end{table}
\end{center}

\begin{center}
\begin{table*}
  \caption{A comprehensive representation of major standardization activities affecting AI Explainability. Note that at the moment of writing [February 2023] for \textit{Publication}, asterisk sign denotes either expected publication release or, if delayed, latest announcement for expected publication release. Note that \textit{INT} in the Area column refers to international standardization organizations, while the column \textit{Delegation} refers to specific committee and/or working groups responsible for standard documents provision.}
  \label{tab:standards}
  \begin{tabular}{p{0.9cm}p{2.5cm}p{2.8cm}p{3.5cm}p{1.7cm}p{1.5cm}}
    \hline
    \textbf{Area} & \textbf{Standardization Body} & \textbf{Delegation} & \textbf{Document(s)} & \textbf{Publication} & \textbf{Release Date} \\
    \hline
    
    \addlinespace[1ex]
    
    EU  & \textbf{CEN - CENELEC} & JTC 21 AI \cite{cencenelec-ai} &  Explainability, Verifiability \cite{cencenelecroadmap}  & Proposal &  TBD \\
    \cline{2-6}

    \addlinespace[2ex]

    & \textbf{ETSI} & ISG SAI \cite{ETSIGRSAI} & DGR/SAI-007 \cite{etsiactivitiesaia, etsi007}  & Under Appr. & 2023-03*\\
    \cline{4-6}
    & & & DGR/SAI-010 \cite{etsiactivitiesaia} & Proposal & TBD \\
    \cline{3-6}

    & & GS ENI & GS ENI 005 v2.1.1 \cite{etsiactivitiesaia, etsieni005}  & Published & 2021 \\
    &&&&&\\
    \hline
    
    \addlinespace[1ex]
     
    US & \textbf{NIST} & NIST Interagency & NISTIR 8312 (Paper) \cite{NIST-8312} & Published & 2021-09 \\
     \cline{4-6}
       &     &  & NISTIR 8367 (Paper) \cite{NIST-8367} & Published & 2021-04 \\
       \cline{4-6}
       &     &  & RMF V1.0 \cite{NIST-AIRMF1.0} & Published & 2023-01 \\
    &&&&&\\
    \hline
    
    \addlinespace[1ex]
    
    UK & \textbf{BSI - NPL} & AI Standard Hub (CDDO, CDEI)
    & Algorithmic Transparency Recording Standard (Guide)
    \cite{UK2021algotransaprencyrecordingstandard}
    & Published & 2021-12 \\\cline{3-6}

    \addlinespace[1ex]
       
    &  & CDDO, Cabinet Office, Office for AI & Ethics, Transparency and Acc. Framework for ADM (Guide) \cite{UK2021EthicsTransparency} & Published & 2021-05\\
    &&&&&\\
    \hline
    
    \addlinespace[1ex]
    
    INT & \textbf{ISO/IEC} & JTC 1/SC 42 AI
    \cite{ISOIECJTC1SC42}
    & ISO/IEC AWI 12792 
    \cite{ISOIECAWI12792}
    & Under Dev. & 2025-02 \\
    &  &  &ISO/IEC AWI TS 6254
    \cite{ISOIECAWITS6254}
    & Under Dev. & 2024-02 \\
    
    &              &     & 
    ISO/IEC TR 24028:2020 
    \cite{ISOIECTRR240282020}
    & Published & 2020-05\\
    \cline{2-6}

    \addlinespace[1ex]
    
    & \textbf{IEEE} & CIS/SC/XAI WG 
    &  P2976 
    \cite{IEEEP2976}
    & Inititation & 2024-07* \\
    \cline{3-6}

    & & VT/ITS 
    & 7001-2021
    \cite{IEEE70012021}
    & Published & 2022-03 \\
    \cline{3-6}

    & & C/AISC/XAI & P2894 (Guide)
    \cite{IEEEP2894}
    & Under Dev. & 2022-03*\\ 
    \cline{3-6}

    & & C/S2ESC/ALGB-WG & P7003
     \cite{IEEEP7003}
    & Published & 2017-02\\ 
    \hline

  \end{tabular}
  
\end{table*}
\end{center}

\end{document}